# On the Parameterised Intractability of Monadic Second-Order Logic

Stephan Kreutzer [⋆]

Oxford University Computing Laboratory, kreutzer@comlab.ox.ac.uk

**Abstract.** One of Courcelle's celebrated results states that if $\mathcal{C}$ is a class of graphs of bounded tree-width, then model-checking for monadic second order logic is fixed-parameter tractable on $\mathcal{C}$ by linear time parameterised algorithms. An immediate question is whether this is best possible or whether the result can be extended to classes of unbounded tree-width.

In this paper we show that in terms of tree-width, the theorem can not be extended much further. More specifically, we show that if $\mathcal{C}$ is a class of graphs which is closed under colourings and satisfies certain constructibility conditions such that the tree-width of $\mathcal{C}$ is not bounded by $\log^{16} n$ then $\text{MSO}_2$-model checking is not fixed-parameter tractable unless SAT can be solved in sub-exponential time. If the tree-width of $\mathcal{C}$ is not poly-logarithmically bounded, then $\text{MSO}_2$-model checking is not fixed-parameter tractable unless all problems in the polynomial-time hierarchy, and hence in particular all problems in NP, can be solved in sub-exponential time.

## 1 Introduction

In 1990, Courcelle proved a fundamental result stating that every property of graphs definable in *monadic second-order logic with edge set quantification* ($\text{MSO}_2$) can be decided in linear time on any class $\mathcal{C}$ of graphs of bounded tree-width. Courcelle's theorem has important consequences both in logic and in algorithm theory. In the design of efficient algorithms on graphs, it can often be used as a simple way of establishing that a property can be solved in linear time on graph classes of bounded tree-width. This is, for instance, useful in relation to algorithms based on tree-width reduction, where input graphs of large tree-width are iteratively reduced until they have small tree-width where the problem is then solved. Often the reduction is the complicated step and it is nice to have a simple way of proving formally that the problem is easy on small-tree-width graphs. Besides being of interest for specific algorithmic problems, results such as Courcelle's and similar *algorithmic meta-theorems* lead to a better understanding how far certain algorithmic techniques, dynamic programming and decomposition in the case of $\text{MSO}_2$, range. See [14,17] for recent surveys on algorithmic meta-theorems.

From a logical perspective, Courcelle's theorem establishes a sufficient condition for tractability of monadic second-order formula evaluation on classes of graphs or

[⋆] Research partly supported by DFG grant KR 2898/1-3. Part of this work was done while the author participated at the workshop "Graph Minors" at BIRS (Banff International Research Stations) in October 2008.

structures: whatever the class $\mathcal{C}$ may look like, if it has bounded tree-width, then $\text{MSO}_2$-model checking is tractable on $\mathcal{C}$.

An obvious question is to ask how tight Courcelle's theorem is, i.e. whether it can be extended to classes of unbounded tree-width and if so, how large the tree-width of graphs in the class can be in general. This is the main motivation for the work reported in this paper.

Given the considerable interest in Courcelle's theorem, it is somewhat surprising that not much is known about limits for $\text{MSO}_2$-model checking, i.e. sufficient conditions for graph classes on which $\text{MSO}_2$-model checking is *intractable*. The question has informally been raised in the community and has led, e.g., to a conjecture by Grohe [14, Conjecture 8.3] that MSO-model checking is not fixed-parameter tractable on any class $\mathcal{C}$ of graphs which is closed under taking subgraphs and whose tree-width is not poly-logarithmically bounded, i.e. there is no constant $c$ such that $\text{tw}(G) \leq \log^c |G|$ for all $G \in \mathcal{C}$. But to the best of my knowledge, the question has so far not been studied systematically.

It follows from the NP-completeness of 3-colourability on planar graphs [12] that MSO-model checking is not fixed-parameter tractable on the class of planar graphs (unless $P = \text{NP}$). Furthermore, it is a simple consequence of the excluded grid theorem that on minor- or topological-minor closed classes of graphs of unbounded tree-width, MSO-model checking is not fixed-parameter tractable unless $P = \text{NP}$ (see Section 2).

In this paper we establish a significantly stronger intractability result by showing that in terms of tree-width, Courcelle's theorem can not be extended much further. Throughout the paper, we will work with coloured graphs, i.e. we will fix a set $\Gamma$ of edge and vertex colours. A class $\mathcal{C}$ of graphs is said to be closed under $\Gamma$-colourings if whenever $G \in \mathcal{C}$ and $G'$ is obtained from $G$ by recolouring, i.e. the underlying undirected graphs are isomorphic, then $G' \in \mathcal{C}$. We will mostly consider classes of graphs closed under colourings. An alternative characterisation is to consider relational structures over a signature $\sigma$ with at most binary relation symbols. We can then fix a class $\mathcal{C}'$ of graphs and consider the class of all finite $\sigma$-structures whose Gaifman-graphs are in $\mathcal{C}$. However, in this paper we prefer to work with coloured graphs rather than Gaifman-graphs of structures. Given a class $\mathcal{C}$ of graphs, we write $\text{MC}(\text{MSO}_2, \mathcal{C})$ for the model-checking problem for $\text{MSO}_2$ on $\mathcal{C}$. The following is the main result of the paper.

**Theorem.** *Let $\Gamma$ be a set of colours with at least one edge and two vertex colours and let $\mathcal{C}$ be a rich and constructible class of $\Gamma$-coloured graphs closed under colourings.*

1. *If the tree-width of $\mathcal{C}$ is not poly-logarithmically bounded, then $\text{MC}(\text{MSO}_2, \mathcal{C})$ is not in XP and hence not fixed-parameter tractable unless all problems in NP (in fact, all problems in the polynomial-time hierarchy) can be solved in sub-exponential time.*
2. *If the tree-width of $\mathcal{C}$ is not bounded by $\log^{16} n$, then $\text{MC}(\text{MSO}_2, \mathcal{C})$ is not in XP and hence not fixed-parameter tractable unless SAT can be solved in sub-exponential time.*

Here, *fixed-parameter tractable* (fpt) means that there is an algorithm which, given $G \in \mathcal{C}$ and $\varphi \in \text{MSO}_2$ decides $G \models \varphi$ in time $f(|\varphi|) \cdot |G|^c$, for some computable function $f$ and constant $c \in \mathbb{N}$. The problem is in XP if it can be solved in time $|G|^{f(|\varphi|)}$. FPT in

the parameterised world corresponds to polynomial-time in the classical framework as the class of problems that can be solved efficiently. XP can be seen as the parameterised exponential-time and is obviously a much larger class of problems than FPT. See [9] for background on parameterised complexity.

Let $\mathcal{C}$ be a class whose tree-width is not bounded by $\log^c n$, for some $c \geq 0$. We call $\mathcal{C}$ *rich for c*, if there is a polynomial $p(x)$ such that for each $n > 0$ there is a graph $G \in \mathcal{C}$ of tree-width $\mathcal{O}(p(n))$ whose tree-width is not bounded by $\log^c |G|$. Essentially, this means that there are not too big gaps between the tree-width of graphs witnessing that the tree-width of $\mathcal{C}$ is not bounded by $\log^c n$. A class of graphs whose tree-width is not bounded by $\log^c n$ is called *rich* if it is rich for $c$. Thus, a class of graphs whose tree-width is not poly-logarithmically bounded is called *rich*, if it is rich for all $c \geq 0$. We will always work with rich classes in this paper for the following reason: take a class of graphs of bounded tree-width and add just one clique. Then the tree-width of this class is not poly-logarithmically bounded but for model-checking purposes this is still essentially a class of bounded tree-width and hence $\text{MSO}_2$-model checking is fixed-parameter tractable on this class. Requiring the classes we work with to be rich rules out such cases. We refer to the following sections for a precise definition of constructible classes but will explain the concept informally below.

Let us give some applications of the theorem. For $c > 0$ let $\mathcal{C}_c$ be the class of all graphs $G$ of tree-width at most $\log^c |G|$. This class is constructible and rich and hence its closure under colourings has intractable $\text{MSO}_2$ model-checking, if $c > 16$. Similarly, the class of planar graphs of tree-width at most $\log^c n$ is rich and constructible. Finally, all (topological) minor-closed classes of unbounded tree-width are constructible and rich. All these examples show that Courcelle's theorem can not be extended to classes of graphs with only poly-logarithmic or a $\log^c n$ bound on the tree-width, for $c > 16$.

**High level description of the proof.** Let us give an intuitive account of the proof of the previous theorem. Clearly, with today's methods we cannot hope to prove that $\text{MSO}_2$-model-checking is fixed-parameter intractable for a class of graphs without relating it to assumptions in computational complexity theory. Consequently, we prove that $\text{MC}(\text{MSO}_2, \mathcal{C})$ is fixed-parameter intractable for a rich and constructible class $\mathcal{C}$ by reducing an NP-complete problem $P$ to $\text{MC}(\text{MSO}_2, \mathcal{C})$ such that if there is an fpt-algorithm for $\text{MC}(\text{MSO}_2, \mathcal{C})$, then $P$ can be solved in sub-exponential time $2^{o(n)}$. More precisely, for each language $P \in \text{NP}$ we first construct a formula $\varphi_P$ and then, given a word $w$ for which we want to test $w \in P$, we construct a graph $G_w \in \mathcal{C}$ such that $G_w \models \varphi_P$ if, and only if, $w \in P$. We will see that the order (number of vertices) of $G_w$ can be bounded by $2^{|w|^{\frac{1}{y}}}$, for some $y > 1$, so that if there was an fpt-algorithm for $\text{MC}(\text{MSO}_2, \mathcal{C})$ with running time $f(|\varphi|) \cdot |G|^c$ then this would imply that $w \in P$ could be decided in time $2^{c|w|^{\frac{1}{y}}} = 2^{o(|w|)}$. For this reduction to work, we need some intermediate steps.

It is well-known that $\text{MSO}_2$-model checking is fixed-parameter intractable on the class of coloured grids (see Figure 1 for an illustration of grids), which can essentially be seen as follows: suppose $P$ can be solved by a non-deterministic Turing-machine $M$ in time $n^c$, where $n$ is the length of the input. Given a word $w$ of length $n$, we choose a $(n^c \times n^c)$-grid $G_w$ and label its top-most row by $w$ from left to right. From the Turing-machine $M$ deciding $P$ we can compute an MSO$_2$-formula $\varphi_M$ depending only on $M$ such that $G_w \models \varphi_M$ if, and only if, $w$ is accepted by $M$ and hence $w \in P$. Essentially, the MSO formula uses the grid to guess the computation table of a successful run of $M$ on $w$. Hence, an fpt-algorithm for MSO-model checking on grids yields a polynomial time algorithm for $P$.

Clearly, if we are just given a class of graphs of tree-width not bounded polylogarithmically, then there is no guarantee that it contains any grids. But we will show that we can define grids in graphs of this class by MSO$_2$-formulas. Adapting a recent proof by Reed and Wood, we first show that if $G$ is a graph of tree-width $k$ then it contains a large structure which we call *coloured pseudo-walls*. Pseudo-walls do not actually occur as minors or sub-graphs of $G$ but as topological minors of certain intersection graphs formed by sets of disjoint paths in $G$. However, it turns out that this is enough to define coloured grids in coloured pseudo-walls by MSO$_2$-formulas. It follows that if the tree-width of $G$ is not bounded by $k$ then we can define an $(l \times l)$-grid in $G$ in MSO$_2$, where $l$ is roughly $\sqrt[10]{k}$ (see Theorem 3.6 for details), and this grid can be coloured. We call a class *constructible* if we can construct these pseudo-walls in graphs $G \in \mathcal{C}$ in polynomial time and it is such classes with wich we work in this paper (see Definition 3.12 for details).

The important aspect here is that the size of the grids we define is polynomially related to the tree-width of the graph, in contrast to the grids guaranteed by the excluded grid theorem (see Theorem 3.1), where the tree-width is exponentially larger than the grids we are guaranteed to find. Hence, using pseudo-walls, if the tree-width of a graph is $\log^k n$ then we can define $(l \times l)$-grids in $G$ for $l \cong \log^{\frac{k}{10}} n$ and this is enough for the reduction sketched above to work.

Obtaining sub-exponential time algorithms for problems such as TSP or SAT is an important open problem in the algorithms community and the common assumption is that no such algorithms exists. This has led to the *exponential-time hypothesis* (ETH) which says that there is no such sub-exponential time algorithm for SAT, a hypothesis widely believed in the community.

Let us briefly comment on the restrictions imposed on the classes $\mathcal{C}$ we study here. While every graph of large enough tree-width contains large pseudo-walls, we do not yet know if we can always compute these structures in polynomial time (and hence we impose the additional restriction to constructible classes). It is conceivable that large pseudo-walls can indeed be computed in polynomial time, i.e. that the proof below can be made algorithmic. This would essentially show that all classes are constructible and effectively remove this condition from our main result. We pose this question as an open problem and leave it to future research.

As mentioned above, Grohe conjectured that MSO-model checking is not fixed-parameter tractable on any class of graphs closed under sub-graphs and whose tree-width is not poly-logarithmically bounded. The statement of the conjecture and the main result of this paper are incomparable as I require closure under colourings whereas Grohe does not. On the other hand, the conjecture requires closure under sub-graphs which I do not. Note that while tree-width is preserved by taking sub-graphs, logarithmic tree-width is not, i.e. a graph whose tree-width is bounded by $\log n$ may contain a sub-graph of order $m$ whose tree-width is not bounded by $\log m$. For instance, the class of planar graphs of tree-width $\log n$ is not closed under sub-graphs, as planar graphs contain grids linear in the tree-width of the graph and these grids can have very high tree-width compared to the number of vertices they contain. So closure under sub-graphs does rule out natural examples of graph classes. On the other hand, Grohe's conjecture does not require colours or constructibility conditions and refers to classes of plain graphs.

Note that it is important for our results that we work with $MSO_2$ and allow quantification over sets of edges as well as over sets of vertices. If we only consider vertex set quantification, i.e. deal with $MSO_1$, then the theorem is false, as for instance, $MSO_1$-model checking is fixed-parameter tractable on the class of cliques, or more generally, on all classes of graphs of bounded clique-width [5].

Following Courcelle's theorem, a series of algorithmic meta-theorems for first-order logic on planar graphs [11], (locally) minor-free graphs [10,6] and various other classes have been obtained. Again, no deep lower bounds, i.e. intractability conditions, are known (see [17] for some simple bounds and [14,17] for recent surveys of the topic). The aim of this paper is to initiate a thorough study of sufficient conditions for intractability in terms of structural properties of input instances.

**Organisation.** We recall monadic second-order logic and what we need about its parameterised complexity in Section 2. The main result is then proved as follows. To show that $MC(MSO_2, \mathcal{C})$ is hard on classes $\mathcal{C}$ of tree-width not poly-logarithmically bounded, we first use a result by Reed and Wood [20] to show that any graph of large enough tree-width contains a structure that is grid-like enough for our purposes. This is proved in Section 3. While these structures do not exist as minors in the graphs, they turn out to be $MSO_2$-definable, which is is shown in Section 4. To use the $MSO_2$-definability for our result, we introduce a new kind of interpretations between structures, called $MSO_2-MSO_2$-*transductions* (see Section 5). Finally, in Section 6, we combine all this to show the main result of the paper.

**Acknowledgements.** I want to thank Mark Weyer for pointing out that the result proved here readily extends to problems in the polynomial time hierarchy.

## 2 Complexity of Monadic Second-Order Logic

We first need some notation and a few concepts from graph theory. For $k \geq 1$, we define $[k] := \{1\ldots,k\}$. We refer to [7] for background on graphs. All graphs are finite and undirected. We write $V(G)$ for the set of vertices and $E(G)$ for the set of edges in a graph $G$. A graph $H$ is a *sub-division* of $G$ (a 1-*subdivision*) if $H$ is obtained from $G$ by replacing edges in $G$ by paths of arbitrary length (of length 2, resp.). $H$ is a *topological minor* of $G$ if a subgraph $G' \subseteq G$ is isomorphic to a sub-division of $H$.

An *elementary wall* $W$ is a graph as in Figure 1 $b$). The cycles of minimal length in $W$ are called *bricks*. A *wall* is a subdivision of an elementary wall. The *height* of a wall is the number of rows of bricks and its *width* the number of columns of bricks. An $l \times k$-wall is a wall of height $l$ and width $k$ and a wall of *order* $l$ is an $l \times l$-wall. Finally, the *nails* of a wall are the vertices of degree 3 in it together with the 4 corners. Hence, in an elementary wall all vertices are nails whereas in a general wall only the vertices of the underlying elementary wall are nails.

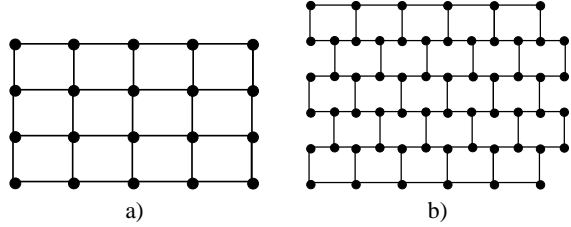

**Fig. 1.** a) A $(4 \times 5)$-grid and b) an elementary wall of order 5.

For the purpose of this paper, it might be easier to think of $k \times k$-grids instead of $k \times k$-walls and everything would go through with grids also. The important property of walls is that their maximum degree is 3. And if a graph $H$ of degree $\leq 3$ is a minor of $G$, then $H$ is also a topological minor of $G$ (see [7, Prop. 1.7.2]). Hence, a sub-division of $H$ actually occurs as sub-graph of $G$. Defining topological minors in MSO$_2$ is much easier than defining minors as we do not need contraction. We will therefore work with walls instead of grids in this paper.

I assume familiarity with basic notions of mathematical logic (see e.g. [8]). In this paper we will only consider signatures $\sigma := \{E, B_1, \ldots, B_s, C_1, \ldots, C_t\}$ of coloured graphs, where $E$ denotes the edge relation, $B_i$ the colours of edges and $C_i$ the colours of graphs. We allow multiple colours per edge or vertex. We denote $\sigma$-structures by Roman letters $A, B, \ldots$. If $R \in \sigma$ is a relation symbol and $A$ a $\sigma$-structure, we write $R(A)$ for the interpretation of $R$ in $A$.

The class of formulas of *monadic second-order logic with edge set quantification* over a signature $\sigma$, denoted MSO$_2[\sigma]$, is defined by the rules for first-order logic with the following additional rules: if $X$ is a second-order variable either ranging over sets of vertices or over sets of edges and $\varphi \in $ MSO$_2[\sigma \dot\cup \{X\}]$, then $\exists X \varphi \in $ MSO$_2[\sigma]$ and $\forall X \varphi \in $ MSO$_2[\sigma]$ with the obvious semantics where, e.g., a formula $\exists F \varphi$, $F$ being a variable over sets of edges, is true in a structure $A$ if there is a subset $F \subseteq E(A)$ such that $(A, F) \models \varphi$. If $\varphi(x)$ is a formula with a free variable $x$ and $A$ is a structure, we write $\varphi(A)$ for the set $\{a : A \models \varphi[a]\}$. See [18] for more on MSO.

In [26], Vardi proved that the model-checking problem MC(MSO$_2$) for MSO$_2$ is PSPACE-complete on the class of all graphs. The complexity of model-checking problems can elegantly be studied in the framework of *parameterised complexity* (see [9] for background on parameterised complexity). If $\mathcal{C}$ is a class of $\sigma$-structures, we define MC(MSO$_2$, $\mathcal{C}$), the *parameterised model-checking problem* for MSO$_2$ on $\mathcal{C}$, as the problem to decide, given $G \in \mathcal{C}$ and $\varphi \in $ MSO$_2[\sigma]$, if $G \models \varphi$. The *parameter* is $|\varphi|$. MC(MSO$_2$, $\mathcal{C}$) is *fixed-parameter tractable* (fpt), if for all $G \in \mathcal{C}$ and $\varphi \in $ MSO$_2[\sigma]$, $G \models \varphi$ can be decided in time $f(|\varphi|) \cdot |G|^k$, for some computable function $f$ and $k \in \mathbb{N}$. The problem is in the class XP, if it can be decided in time $|G|^{f(|\varphi|)}$.

As, for instance, the NP-complete problem 3-Colourability is definable in MSO$_2$, MC(MSO$_2$, GRAPHS), the model-checking problem for MSO$_2$ on the class of all graphs, is not fixed-parameter tractable unless $P = $ NP. However, if we restrict the class of graphs then we can obtain much better results.

For instance, Courcelle proved that MSO$_2$-model-checking is fixed-parameter tractable on graph classes of bounded tree-width. Tree-width is a global connectivity measure of graphs that was introduced by Robertson and Seymour in their graph minor series. We will work with the dual notion of brambles below and therefore refer the reader to [7] for a definition of tree-width.

**Definition 2.1.** Let $f : \mathbb{N} \to \mathbb{N}$ be a function and $\mathcal{C}$ be a class of graphs. The tree-width of $\mathcal{C}$ is *bounded by $f$*, if $\operatorname{tw}(G) \leq f(|G|)$ for all $G \in \mathcal{C}$. $\mathcal{C}$ has *bounded tree-width* if its tree-width is bounded by a constant.

Many natural classes of graphs, for instance series-parallel graphs, are found to have bounded tree-width.

**Theorem 2.2** (*Courcelle* [4]). *MC(MSO$_2$, $\mathcal{C}$) is fixed-parameter tractable on any class $\mathcal{C}$ of graphs of tree-width bounded by a constant.*

Courcelle's theorem gives a sufficient condition for MC(MSO$_2$, $\mathcal{C}$) to be tractable. The obvious counterpart are sufficient conditions for intractability, i.e. what makes MSO$_2$-model checking hard? Surprisingly, it seems that so far – in terms of conditions imposed on the structures – this question has not received much attention. Garey, Johnson and Stockmeyer [12] proved that 3-Colourability remains NP-hard on the class of planar graphs of degree at most $4$. It follows that unless $P = \operatorname{NP}$, MC(MSO$_2$, PLANAR) is not fixed-parameter tractable, where PLANAR denotes the class of planar graphs. We mention a few other intractability results which will be useful later in the paper. We first need the following lemma whose proof is folklore.

**Lemma 2.3.** *Let $M$ be a non-deterministic Turing-machine. There is a formula $\varphi_M \in$ MSO$_2$ such that for all words $w \in \Sigma^*$, if $G$ is a $k \times k$-wall whose top-most row is coloured by $w$ from the left, then $G \models \varphi_M$ if, and only if, $M$ accepts $w$ in at most $k$ steps. Furthermore, the formula $\varphi_M$ can be constructed effectively from $M$. The same holds if $M$ is an alternating Turing-machine with a bounded number of alternations, as they are used to define the polynomial-time hierarchy.*

The proof uses the idea that we can simply guess the computation table of $M$ running on $w$ on the wall using set quantification allowed in MSO$_2$. The following lemma shows that MSO$_2$-model checking is not fixed-parameter tractable on the class of walls.

**Lemma 2.4.** *Let $\mathcal{C}$ be the class of walls. Then MC(MSO$_2$, $\mathcal{C}$) is not in XP, and hence not fpt, unless* EXPTIME $=$ NEXPTIME.

*Proof.* Consider the following problem. Given a non-deterministic Turing machine $M$ and a number $t$ encoded in unary, the problem is to decide if $M$ has an accepting run on the empty word of length exactly $t$. The parameter is $|M|$. In [1], Aumann and Dombb proved that this problem is not in XP unless EXPTIME $=$ NEXPTIME. See also [3].

Given $M$ and $t$ we can construct a formula $\vartheta$ as in Lemma 2.3 but which also checks that all rows of the grid are being used, i.e. that the computation on a $t \times t$-grid $G_{t \times t}$ makes $t$ steps. Then $G_{t \times t} \models \vartheta$ if, and only if, $M$ has an accepting run of length $t$ on the empty input. □

Lemma 2.3 combined with the excluded grid theorem 3.1 below establish the following result

**Lemma 2.5.** *If $\mathcal{C}$ is closed under minors and has unbounded tree-width, then* $\text{MC}(\text{MSO}_2, \mathcal{C})$ *is not fixed-parameter tractable unless* $P = \text{NP}$. *This can be strengthened to closure under topological minors.*

*Proof.* By the excluded grid theorem, if $\mathcal{C}$ is closed under minors and has unbounded tree-width, then it contains all $(k \times k)$-grids.

Let $P$ be an NP-complete problem over the alphabet $\Sigma := \{0, 1\}$ decidable by a non-deterministic Turing-machine $M$ in time $n^c$, for some $c \in \mathbb{N}$. By Lemma 2.3, we can compute an $\text{MSO}_2$-formula $\varphi_M$ such that for all words $w \in \Sigma^*$, if $G_w$ is a $(k \times k)$-grid, for $k = |w|^c$, whose top-most row is coloured by $w$ from the left, then $M$ accepts $w$ if, and only if, $G_w \models \varphi_M$. Now, the grids in $\mathcal{C}$ are not coloured but we can easily simulate the labelling of the top row of the grid by deleting edges as follows: let $w := w_1, \ldots, w_l$ and $G_w$ be a $(2k \times 2k)$-grid, where $k$ is as before, where we delete in the top row from left the edge between vertex $2i$ and $2i + 1$ if $w_i = 1$. Now we can easily modify the formula $\varphi_M$ so that it decodes the word $w$ from this encoding in an uncoloured grid. The result then follows by the same proof as for Lemma 2.3.

The second part follows from the fact that every graph of large enough tree-width contains a $k \times k$-wall as a topological minor and that walls can be used in the same way as grids to show fixed-parameter intractability for MSO. □

## 3 Pseudo-Walls in Graphs

One of the fundamental results of Robertson and Seymour's theory of graph minors is the excluded grid theorem [22], saying that there is a computable function $f : \mathbb{N} \to \mathbb{N}$ such that every graph of tree-width at least $f(k)$ contains a $k \times k$-grid as a minor. The best explicit bound known for the function $f$ is given by the following theorem.

**Theorem 3.1** (*Robertson, Seymour, Thomas [21]*). *Every graph of tree-width at least $20^{2 \cdot k^5}$ contains a $k \times k$ grid as a minor.*

Robertson et al. [21] also proved that there are graphs of tree-width proportional to $k^2 \log k$ that do not contain $G_{k \times k}$ as a minor. So far this is the best lower bound known for the function $f$ above. In particular it is open whether polynomial tree-width forces large grid-minors.

In [20] Reed and Wood consider a different type of obstructions to small tree-width, called *grid-like* minors. A grid-like minor of order $l$ in a graph $G$ is a set $\mathcal{P}$ of paths such that the intersection graph $\mathcal{I}(\mathcal{P})$ contains a $K_l$-minor, where $K_l$ denotes the complete graph on $l$ vertices. Here, the *intersection graph* of a set $\mathcal{P}$ of paths is the graph with vertex set $\mathcal{P}$ and an edge between two paths $P, Q \in \mathcal{P}$ if $P \cap Q \neq \emptyset$. If $\mathcal{P}, \mathcal{Q}$ are sets of paths in $G$, we write $\mathcal{I}(\mathcal{P}, \mathcal{Q})$ for $\mathcal{I}(\mathcal{P} \dot\cup \mathcal{Q})$, the intersection graph formed by their disjoint union. The main result of their paper is to show that polynomial tree-width forces large grid-like minors.

**Theorem 3.2** (*Reed, Wood [20]*). *Every graph of tree-width at least $ck^4 \sqrt{\log k}$ contains a grid-like minor of order $k$, for some constant $c$. Conversely, every graph that contains a grid-like minor of order $l$ has tree-width at least $\lceil \frac{l}{2} \rceil - 1$.*

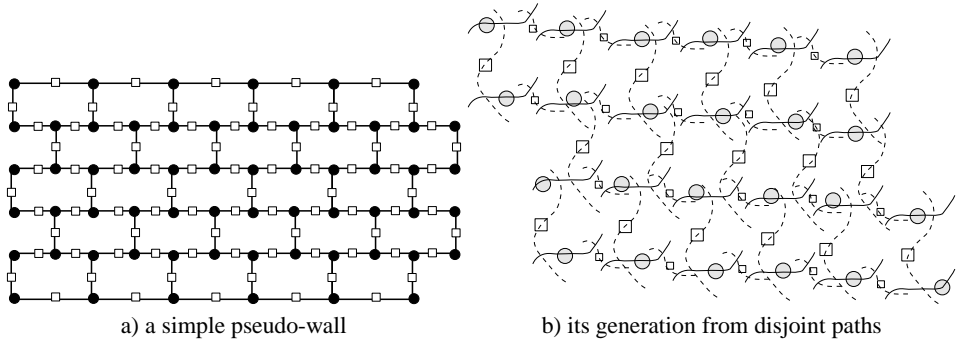

a) a simple pseudo-wall    b) its generation from disjoint paths

**Fig. 2.** A simple pseudo-wall and the paths $\mathcal{P}$ (solid) and $\mathcal{Q}$ (dashed) generating it.

While I do not yet know how to use this result directly, we can use its proof to find the structures in $G$ we need.

**Definition 3.3.** A *pseudo-wall* of order $l$ in $G$ is a pair $(\mathcal{P}, \mathcal{Q})$ of sets of disjoint paths in $G$ such that $\mathcal{I}(\mathcal{P}, \mathcal{Q})$ is a wall of order $l$.

We will see below that every graph of large enough tree-width contains a large pseudo-wall and that these can be defined in MSO$_2$. Essentially, to show that MSO$_2$ model-checking is fixed-parameter intractable on a class $\mathcal{C}$ of large enough tree-width, we will use pseudo-walls in a similar way as grids are used in Lemma 2.3. In particular, we want to label the top-most row of the pseudo-wall by a word $w$ over a finite alphabet. However, pseudo-walls do not occur as subgraphs of the graphs $G$, which makes labelling them somewhat more difficult. Instead, we have to colour the graph $G$ so that this colouring induces the labelling by the word $w$ of the pseudo-wall it contains. The main difficulty is that the colouring of $G$ must induce a unique labelling of the pseudo-wall and that both the pseudo-wall as well as its labelling can be defined inside $G$ by MSO$_2$-formulas. Unfortunately, this makes the definition of a coloured pseudo-wall technically somewhat more complicated, although the idea is still simple. Let $\Sigma$ be a set of colours and let $B$ be an additional colour for edges and $R$ an additional colour for vertices. Let $\Gamma := \{B, R\} \,\dot\cup\, \Sigma$.

**Definition 3.4.** A $\Sigma$-*coloured pseudo-wall* of order $l$ in a $\Gamma$-coloured graph $G$ is a triple $(\mathcal{P}, \mathcal{Q}, A)$ such one of the following holds:

**Simple pseudo-walls.** $\mathcal{I}(\mathcal{P}, \mathcal{Q})$ is a 1-subdivision of an elementary wall $W$ of order $l$ such that the vertices of $W$ (which we called *nails* above) are exactly the paths in $\mathcal{P}$. See Figure 2 for an illustration. Figure 2 a) shows the pseudo-wall, where the solid black circles are the vertices from $\mathcal{P}$ and squares denote the vertices from $\mathcal{Q}$. Figure 2 b) shows how (a part of) this pseudo-wall corresponds to paths in $G$, where dashed lines respresent paths in $\mathcal{Q}$ and solid lines paths in $\mathcal{P}$. Note, though, that in general the paths could intersect in much more complicated ways than displayed.

Let $\mathcal{P} := \{P_1, \ldots, P_k\}$ be such that $P_1 \ldots P_l$ form the nails of the top-most row of $W$ in order from left to right. Recall that each $P_i$ is a path in $G$. Then $A$ is the path in $G$ obtained from the "concatenation" $P_1 \cdot P_2 \cdots P_k$, i.e. $V(A) := \bigcup_{1 \leq i \leq k} V(P_i)$ and $E(A) := \bigcup_{1 \leq i \leq k} E(P_i)$ together with additional edges connection one endpoint of $P_i$ to an endpoint of $P_{i+1}$, for $1 \leq i < k$, so that this results in a path.

Furthermore, the edges in $E(A)$ are coloured $B$. The two endpoints of each $P_i$ are coloured $R$ and the vertices in the paths $P_1, \ldots, P_l$ carry colours from $\Sigma$ so that all vertices in a path $P_i$ are coloured by the same colour from $\Sigma$.

This colouring of $G$ induces a labelling of the wall of order $l$ where the nails $v_1, \ldots, v_l$ in the top-most row are labelled so that $v_i$ is labelled by the colour $C_i \in \Sigma$ of the path $P_i$. If $w := C_1 \cdots C_l$ is the sequence of colours on $P$ we say that $(\mathcal{P}, \mathcal{Q}, A)$ *encodes* the word $w \in \Sigma^*$.

The motivation behind simple pseudo-walls is as follows. If we find this structure in a graph $G$ then the path $A$ tells us what the top-most row of the wall is and it also gives us an order of the top-most row vertices. Colouring $A$ by $B$ will enable us to define this coloured pseudo-wall in MSO$_2$. If we want to encode a word $w := w_1, \ldots, w_l \in \Sigma^*$ in the wall then we can simply label the paths $P_1, \ldots, P_l$ in $G$ which form the top-most row of the wall by $w_1, \ldots, w_l$ and this induces the correct labelling of the wall $\mathcal{I}(\mathcal{Q}, \mathcal{P})$.

**Complex pseudo-walls.** Complex walls are structures as illustrated in Figure 3. Essentially, they consist of a subdivision $W'$ of a wall $W$ in $\mathcal{I}(\mathcal{P}, \mathcal{Q})$. To define the colouring of the wall, there will be additional paths $\mathcal{I}(\mathcal{P}, \mathcal{Q})$ connecting some of the vertices of the top-most row of $W'$ to the path $A$ so that the order is preserved, i.e. the paths do not "cross". We can then colour the path $A$ and thereby induce a colouring of the top-most row.

Formally, for complex coloured pseudo-walls, $A$ is a path in $G$ such that each $U \in \mathcal{P}$ has exactly one endpoint in $A$ and no path in $\mathcal{Q}$ has an endpoint in $A$. Furthermore, there are subsets $\mathcal{P}' \subseteq \mathcal{P}$ and $\mathcal{Q}' \subseteq \mathcal{Q}$ such that $\mathcal{I}' := \mathcal{I}(\mathcal{P}', \mathcal{Q}')$ is a wall of order $l$.

Let $T \subseteq \mathcal{I}'$ be the top-most row of the wall and let $x_1 \ldots x_k$ be the vertices of $T$ in order from left to right. Let $I := \{i_1, \ldots, i_l\}$ be the index set such that $x_{i_1}$ is the top-left corner, $x_{i_l}$ is the top-right corner and $(x_{i_j})_{1<j<l}$ lists the vertices in $T$ of degree 3 in order from left to right. For $1 < s < t \leq l$ let $T(s,t)$ be the segment of $T$ between $x_{i_s}$ and $x_{i_t}$ including the latter but not the former. We define $T[0,1]$ to be the segment containing the vertices $x_{i_1}, \ldots x_{i_2}$. Now, the sets $\mathcal{P}_r := \mathcal{P} \setminus \mathcal{P}'$ and $\mathcal{Q}_r := \mathcal{Q} \setminus \mathcal{Q}'$ induce disjoint paths $P_1 \ldots P_l$ in $\mathcal{I} := \mathcal{I}(\mathcal{P}, \mathcal{Q})$ (i.e. each $P_i$ consists of a set of paths in $G$) such that

- one endpoint of each path $P_i$ in $\mathcal{I}$ is incident to a vertex $x_i$ of the top-most row of the wall so that each $T(s,t)$, for $1 < s < t \leq l$, contains exactly one $x_i$ and $T[0,1]$ contains 2 and
- for the other endpoint $u_i$ of $P_i$ in $\mathcal{I}(\mathcal{P}, \mathcal{Q})$ (which is a path in $G$) we have $\{v_i\} = u_i \cap A$, where $v_i \in V(G)$, and
- $v_1, \ldots, v_l$ occur in this order on $A$.

Now suppose $v_1 \ldots v_l$ are coloured by $C_1 \ldots C_l$ respectively. Then this colouring induces the labelling of $\mathcal{I}(\mathcal{P}', \mathcal{Q}')$ where $x_{i_s}$ gets colour $C_s$, $1 \leq s \leq l$. We say that $(\mathcal{P}, \mathcal{Q}, A)$ *encodes* the word $w := C_1 \ldots C_l$.

A crucial feature of pseudo-walls in coloured graphs is that they are unique in the sense that if $G$ is a graph coloured by $\{B, R\} \dot{\cup} \Sigma$, then every pseudo-wall $(\mathcal{P}, \mathcal{Q}, A)$ in $G$ encodes the same word $w$ (there may be no coloured pseudo-wall in $G$).

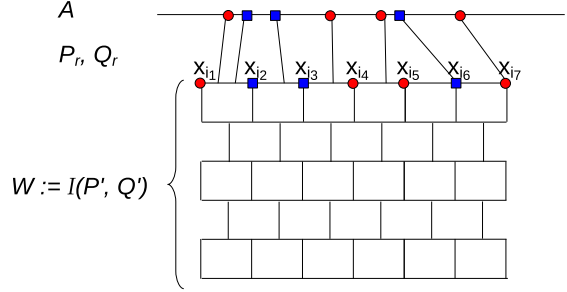

**Fig. 3.** A complex pseudo-wall.

This is obvious for simple pseudo-walls, as the path $A$ is uniquely determined by its colouring ($B$-edges and the $P_i$'s separated by $R$-vertices) and this uniquely determines the colouring of the wall and hence the encoded word. For complex walls, the path $A$ is again determined by its colouring and this fixes the order of the colours occuring on $A$ and hence on the wall. Here we use that the paths connecting $A$ to the wall preserve the order.

**Definition 3.5.** We say that a graph $G$ encodes $w \in \Sigma^*$ if it contains a $\Sigma$-coloured pseudo-wall encoding $w$. We say that $G$ *encodes $w$ with power $k$*, for some $k \geq 1$, if $G$ contains a $\Sigma$-coloured pseudo-wall of order $|w|^k$ encoding $w$.

The proof of the next theorem is essentially the proof of Theorem 3.2 with some modifications to get coloured pseudo-walls instead of grid-like minors.

**Theorem 3.6.** *There is a constant $c$ such that if $G$ is a graph of tree-width at least $c \cdot m^8 \cdot \sqrt{\log(m^2)}$, then $G$ contains a $\Sigma$-coloured pseudo-wall of order $m$.*

We need a few preparations to present the proof. Let $G$ be a graph. Two subgraphs $X, Y \subseteq G$ *touch* if $X \cap Y \neq \varnothing$ or there is an edge in $G$ linking $X$ and $Y$. A *bramble* in $G$ is a set $\mathcal{B}$ of pairwise touching connected subgraphs. A set $S$ is a *cover* for $\mathcal{B}$ if $S \cap B \neq \varnothing$ for all $B \in \mathcal{B}$. The *order* of $\mathcal{B}$ is the minimum size of a cover of $\mathcal{B}$. Brambles provide a dual characterisation of tree-width as shown by the following theorem.

**Theorem 3.7** (*Seymour and Thomas* [23]). *A graph $G$ has treewidth at least $k$ if, and only if, $G$ contains a bramble of order at least $k + 1$.*

We first need the following lemma, whose simple proof is included in the appendix.

**Lemma 3.8** (*Birmelé, Bondy, Reed* [2]). *Let $\mathcal{B}$ be a bramble in a graph $G$. Then $G$ contains a path intersecting every element in $\mathcal{B}$.*

*Proof.* Pick a bramble element $B \in \mathcal{B}$ and choose a vertex $v \in V(B)$. We initialise $P := v$ and maintain the invariant that for one endpoint $v$ of $P$ there is a bramble element $B \in \mathcal{B}$ such that $\{v\} = V(B) \cap V(P)$. While there still is a bramble element $B' \in \mathcal{B}$ not containing a vertex of $P$ choose a path $P'$ from $v$ to $B'$ in $B \cup B'$ as short as possible. Such as path exists as $B$ and $B'$ touch. As $P'$ is choosen as short as possible, all internal vertices are contained in $B$ and one endpoint of $P'$ is the only element of $P'$ in $B'$. In particular, as $\{v\} = V(B) \cap V(P)$ this path $P'$ has no vertex in common with $P$ other than $v$. Further, $P \cup P'$ is still a path satisfying the invariant. We proceed until there are no bramble elements left which have an empty intersection with $P$. □

Clearly, if $P$ is a path in $G$ intersecting every element of a bramble $\mathcal{B}$ then the length of $P$ must be at least the order of $\mathcal{B}$. The following lemma is proved by combining the previous lemma with an application of Menger's theorem. The proof appears in the appendix.

**Lemma 3.9** (*Reed and Wood* [20]). *Let $G$ be a graph containing a bramble $\mathcal{B}$ of order at least $kl$, for some $k, l \geq 1$. Then $G$ contains $l$ disjoint paths $P_1, \ldots, P_l$, and for distinct $i, j \in [l]$, $G$ contains $k$ disjoint paths between $P_i$ and $P_j$.*

*Proof.* By Lemma 3.8, there is a path $P := (v_1 \ldots v_n)$ in $G$ intersecting every element of $\mathcal{B}$ and hence of length at least $kl$. For $1 \leq i \leq j \leq n$ let $P_{ij}$ be the sub-path of $P$ induced by $\{v_i, \ldots, v_j\}$. Let $t_1$ be the minimal integer such that the sub-bramble $\mathcal{B}_1 := \{B \in \mathcal{B} : B \cap P_{ij} \neq \varnothing\}$ has order $k$. Given $t_i, \mathcal{B}_i$ with $i < l$, let $t_{i+1}$ be the minimal integer such that the sub-bramble $\mathcal{B}_{i+1} := \{B \in \mathcal{B} : B \cap P_{t_i t_{i+1}} \neq \varnothing, B \cap P_{1 t_i} = \varnothing\}$ has order $k$. Since $\mathcal{B}$ has order $kl$ we obtain in this way integers $t_1 < t_2 < \cdots < t_l \leq n$. Let $P_i := P_{t_{i-1}+1, t_i}$, where $t_0 := 0$. By construction, the $P_i$ are pairwise disjoint.

Suppose there is a set $S \subseteq V(G)$ of cardinality $|S| < k$ separating some $P_i$ and $P_j$. Hence, $S$ is neither a hitting set of $\mathcal{B}_i$ nor of $\mathcal{B}_j$ and hence there is $B_i \in \mathcal{B}_i$ and $B_j \in \mathcal{B}_j$ such that $S \cap (B_i \cup B_j) = \varnothing$. As $B_i$ and $B_j$ touch it follows that $S$ does not separate $B_i$ and $B_j$ and therefore does not separate $P_i$ and $P_j$. Hence, any set separating $P_i$ and $P_j$ must be of cardinality at least $k$. By Menger's theorem, there are $k$ disjoint paths between $P_i$ and $P_j$ as required. □

Recall that a graph $G$ is $d$-*degenerated* if every subgraph of $G$ contains a vertex of degree at most $d$. Mader [19] proved that every graph with no $K_l$-minor is $2^{l-2}$-degenerated. Let $d(l)$ be the minimal integer such that every graph with no $K_l$-minor is $d(l)$-degenerated. Kostocha and, independently, Thomason showed that $d(l) \in \Theta(l\sqrt{\log l})$.

**Theorem 3.10** (*Kostocha*[16]; *Thomason* [24,25]). *Every graph with no $K_l$-minor is $d(l)$-degenerated, where $d(l) \leq cl\sqrt{\log l}$ for some constant $c$.*

We are now ready to prove Theorem 3.6.

*Proof of Theorem 3.6.* Let $l := m^2$. Hence, $\operatorname{tw}(G) \geq c \cdot l^4 \cdot \sqrt{\log l}$. Let $k := \lceil 2e(2\binom{l}{2} - 3)d(l) \rceil$. Choose $c$ so that $cl^4\sqrt{\log l}$ is at least $kl - 1$ and let $G$ be a graph of treewidth at least $cl^4\sqrt{\log l}$. By Theorem 3.7, $G$ contains a bramble of order at least $kl$ and therefore, by Lemma 3.9, $G$ contains a path $P$ that can be decomposed into $l$ disjoint paths $P_1, \ldots, P_l$ and, for distinct $i, j \in [l]$, $G$ contains a set $\mathcal{Q}_{i,j}$ of $k$ disjoint paths between $P_i$ and $P_j$.

For distinct $i, j \in [l]$ and $a, b \in [l]$ such that $\{i, j\} \neq \{a, b\}$, let $H_{i,j,a,b} := \mathcal{I}(\mathcal{Q}_{i,j}, \mathcal{Q}_{a,b})$ be the intersection graph of $\mathcal{Q}_{i,j} \cup \mathcal{Q}_{a,b}$.

Suppose there are $i, j, a, b$ as above such that $K_l \preceq H_{i,j,a,b}$. We define a complex pseudo-wall $(\mathcal{P}, \mathcal{Q}, A)$ as follows. Set $A := P_i$. Let $X_1, \ldots, X_l$ be the connected subgraphs in $H := H_{i,j,a,b}$ constituting the image of $K_l$ in $H$. W.l.o.g. we assume that each $X_i$ contains a path from $\mathcal{P}$. (There can only be at most one $X_i$ consisting of a single path from $\mathcal{Q}$.) Fix a direction of $A$, i.e. define one endpoint to be "smaller" than the other. This direction induces an odering $\sqsubset$ on the paths in $\mathcal{P}$ where $P \sqsubset P'$ if $V(P) \cap V(A)$ is smaller than $V(P') \cap V(A)$. We order the sets $X_i$ by letting $X_i < X_j$ if there is a vertex $P \in X_i \cap \mathcal{P}$ such that $P \sqsubset P'$ for all $P' \in X_j \cap \mathcal{P}$. W.l.o.g. we assume that $X_1 < X_2 < \cdots < X_l$. We will take $X_1, \ldots, X_{\sqrt{l}}$ to form the top-most row of the wall $W$ in order from left to right. As $X_1, \ldots, X_l$ constitute a $K_l$ minor, we can choose in each $X_i$ a tree $T_i \subseteq X_i$ with at most three leaves such that $\bigcup_{i \leq l} T_i$ together with suitable edges joining distinct trees form a wall $W$ of order $\sqrt{l}$ with $T_1 \ldots T_{\sqrt{l}}$ forming the top-most row, i.e. the vertices of degree 3 in $T_2, \ldots, T_{\sqrt{l}-1}$ and a vertex from each of $T_1$ and $T_{\sqrt{l}}$ form the nails of the top-most row. We set $\mathcal{P}' := V(W) \cap \mathcal{P}$ and $\mathcal{Q}' := V(W) \cap \mathcal{Q}$. Now, for $1 \leq i \leq \sqrt{l}$ pick a path $P_i$ in $X_i$ from the "nail" in $T_i$ to the $\sqsubset$-minimal vertex $P_i^m \in V(X_i) \cap \mathcal{P}$ and let $\mathcal{Z}$ be the union of these paths, which obviously are disjoint. We set $\mathcal{P}_r := \mathcal{Z} \cap \mathcal{P}$ and $\mathcal{Q}_r := \mathcal{Z} \cap \mathcal{Q}$. Then $(\mathcal{P}' \cup \mathcal{P}_r, \mathcal{Q}' \cup \mathcal{Q}_r, A)$ form a complex pseudo-wall of order $l$.

Now suppose that $K_l \not\preceq H_{i,j,a,b}$ for all $i, j, a, b$ as above. Then, by Theorem 3.10, all $H_{i,j,a,b}$ are $d(l)$-degenerated. Let $H$ be the intersection graph of $\bigcup\{\mathcal{Q}_{i,j} : 1 \leq i < j \leq l\}$. Obviously, $H$ is $\binom{k}{2}$-colourable with each $\mathcal{Q}_{i,j}$ being a colour class. Each colour class has $k$ vertices and each pair of colour classes induce a $d(l)$-degenerated graph. The following lemma is from [20].

**Lemma 3.11.** *Let $r \geq 2$ and let $V_1, \ldots, V_r$ be the colour classes in an $r$-colouring of a graph $H$. Suppose that $|V_i| \geq n := 2e(2r - 3)d$, for all $i \in [r]$, and $H[V_i \cup V_j]$ is $d$-degenerated for distinct $i, j \in [r]$. Then there exists an independent set $\{x_1, \ldots, x_r\}$ of $H$ such that each $x_i \in V_i$.*

Applying the lemma to our setting, with $n = k$ and $r = \binom{l}{2}$ and $d = d(l)$, we obtain an independent set in $H$ with one vertex in each colour class. That is, in each set $\mathcal{Q}_{i,j}$ there is one path $Q_{i,j}$ such that $Q_{i,j} \cap Q_{a,b} = \varnothing$. Hence, there is a subset $\mathcal{Q}$ of $\{Q_{i,j} : i \neq j\}$ so that $\mathcal{I}(\mathcal{P}, \mathcal{Q})$ is the 1-subdivision of a wall of order $l$. Then, $(\mathcal{P}, \mathcal{Q}, P)$ form a simple pseudo-wall of order $l$. As $m := l^2$, we either obtain a pseudo-wall (simple or a complex) of order $m$. □

We can now give a formal definition of *constructible* and *rich* classes of graphs.

**Definition 3.12.** Let $\mathcal{C}$ be a class of graphs closed under $\Gamma$-colourings.

1. $\mathcal{C}$ is called *constructible* if in every graph $G \in \mathcal{C}$ of tree-width at least $c \cdot m^8 \cdot \sqrt{\log(m^2)}$, where $c$ is from Theorem 3.6, we can compute in polynomial time a coloured pseudo-wall of order $m$.
2. Let $c \in \mathbb{N}$. We call $\mathcal{C}$ *rich for $c$*, if there is a polynomial $p(x)$ such that for each $n > 0$ we can compute in polynomial time (in the size of $G$) a graph $G \in \mathcal{C}$ of tree-width $\mathcal{O}(p(n))$ whose tree-width is not bounded by $\log^c |G|$. A class of graphs whose tree-width is not bounded by $\log^c n$ is called *rich* if it is rich for $c$. Thus, a class of graphs whose tree-width is not poly-logarithmically bounded is called *rich*, if it is rich for all $c \geq 0$.

It is conceivable that the large pseudo-walls whose existence we proved above can always be computed in polynomial time, i.e. that the proof can be made algorithmic. This would imply that all classes of graphs are constructible. We leave this for future research.

However, various natural classes of graphs are easily seen to be constructible. Examples include classes of planar graphs, as they contain large walls as sub-graphs and the rows and columns of a wall yield a pseudo-wall, the class of all graphs of tree-width at most $\log^c n$, for some $c$ and every class of graphs of tree-width not bounded by a constant which is closed under minors or topological-minors. Finally, richness and constructibility is preserved by taking super-classes, i.e. if $\mathcal{C} \subseteq \mathcal{C}'$ and $\mathcal{C}$ is constructible or rich, then so is $\mathcal{C}'$.

The theorem shows that in any graph of sufficiently large tree-width we find a large pseudo-wall. We will show below that this is enough to define large walls in graphs of large tree-width by means of MSO$_2$-formulas. It follows from Theorem 3.6 above that if $\mathcal{C}$ is a class of graphs of unbounded tree-width which is closed under colouring then for each $w \in \Sigma^*$, $\mathcal{C}$ contains a graph encoding $w$. In fact, for each $c \geq 1$, $G$ contains a graph encoding $w$ with power $c$. The following lemma summarises what we will need about colourings in the following sections.

**Lemma 3.13.** *Let $\mathcal{C}$ be a class of graphs closed under $\Gamma$-colourings and let $w \in \Sigma^*$ be a word of length $m$. If there is a graph $G \in \mathcal{C}$ of tree-width $c \cdot (m^k)^8 \cdot \sqrt{\log(m^k)^2}$, where $c$ is the constant from Theorem 3.6, whose tree-width is not bounded by $\log^{8k} |G|$ then there is a graph $G \in \mathcal{C}$ encoding $w$ with power $k$ such that $|G| < 2^{c' \cdot m^{\frac{1}{y}}}$, for some constants $y > 1$ and $c' := c(k)$ depending on $k$ but not on $w$.*

*Proof.* Let $w$ be given with $|w| = m$. We want to generate a $\Sigma$-coloured pseudo-wall of order $m^k$. By Theorem 3.6, if $\operatorname{tw}(G) = c \cdot (m^k)^8 \cdot \sqrt{\log(m^k)^2}$ then $G$ contains such a pseudo-wall and hence a colouring of $G$ encodes $w$ with power $k$. By the closure conditions of $\mathcal{C}$ we can find such a graph $G \in \mathcal{C}$ such that in addition $\operatorname{tw}(G) \geq \log^x |G|$, for some $x > 8k$.

Hence, $\log^x |G| < c \cdot (m^k)^8 \cdot \sqrt{\log((m^k)^2)}$ iff $\log |G| < \sqrt[x]{c \cdot (m^k)^8 \cdot \sqrt{\log(m^k)^2}}$ iff $\log |G| < c^{\frac{1}{x}} \cdot m^{\frac{8k}{x}} \cdot (2k \log m)^{\frac{1}{2x}}$. This implies $|G| < 2^{c' \cdot m^{\frac{1}{y}}}$, for some $y > 1$ and $c'$. □

## 4 Defining coloured pseudo-walls in graphs of large tree-width

In this section we aim at defining $\Sigma$-coloured pseudo-walls in graphs of large enough tree-width in MSO$_2$. Fix a set $\Sigma$ of colours and let $\Gamma := \Sigma \dot{\cup} \{B, R\}$ be as defined in Section 3. Let $G$ be a $\Gamma$-coloured graph and $\mathcal{P}, \mathcal{Q}, A \subseteq E(G)$ be sets of edges. For $(\mathcal{P}, \mathcal{Q}, A)$ to be a $\Sigma$-coloured pseudo-wall in $G$, we first need to say that $\mathcal{P}$ and $\mathcal{Q}$ are sets of pairwise disjoint paths in $G$. Note that $\mathcal{P}$ induces a set of pairwise disjoint paths if, and only if, $i)$ every vertex $v \in G$ is incident to at most two edges in $\mathcal{P}$ and $ii)$ the subgraph of $G$ induced by the edges in $\mathcal{P}$ is acyclic. This can easily be defined in MSO$_2$ and we will see the formulas below for the more complicated case of paths and acyclicity in $\mathcal{I}(\mathcal{P}, \mathcal{Q})$. Furthermore, we have to say that the edges of $A$, and only those, are coloured by $B$. Now, we have to distinguish between simple and complex coloured pseudo-walls. This can easily be done in MSO$_2$ as in the first case $\bigcup_{P \in \mathcal{P}} P \subseteq A$ (at least in the pseudo-walls generated in the previous section, in general pseudo-walls this is only true for the paths in the top-most row, but that could equally be used to distinguish the two types of walls) whereas this fails in the second. We will present the case for simple pseudo-walls explicitly. The other case follows analogously using the same ideas.

We first need a few auxiliary formulas. To ease the presentation we assume that no path $P$ occurs in both $\mathcal{P}$ and $\mathcal{Q}$. This is guaranteed by the pseudo-walls generated in Section 3 but we could also easily modify the formulas below to avoid this assumption (see also Section 5).

In what follows we will use MSO$_2$-formulas, interpreted in $G$, to speak about the intersection graph $\mathcal{I} := \mathcal{I}(\mathcal{P}, \mathcal{Q})$. To increase readability of formulas we agree on the following convention: lower case letters are used for first-order variables, variables $P, Q, ...$ range over sets of edges and variables $E, F, H$ range over sets of vertices. It may seem bizarre to use $F, H$ for a set of vertices. The reason will become clear below as we will be using variables $E$ for sets of vertices in $G$ which represent sets of edges in $\mathcal{I}$. As a final piece of notation, we write "$P \in \mathcal{P}$" to say that $P$ is a component of $\mathcal{P}$, i.e. one of the paths in $\mathcal{P}$, and analogously for $Q \in \mathcal{Q}$. Furthermore, we will write $x \in V(P)$ for the formula $\exists y P x y$ to say that $x$ is adjacent to an edge in $P$.

Recall that for two paths $P \in \mathcal{P}$ and $Q \in \mathcal{Q}$ there is an edge $\{P, Q\} \in E(\mathcal{I})$ if $P \cap Q \neq \varnothing$ in $G$. Note that, as $\mathcal{P}$ and $\mathcal{Q}$ are sets of disjoint paths, there are no three distinct paths in $\mathcal{P} \cup \mathcal{Q}$ intersecting in a single vertex. Hence, we can represent edges $\{P, Q\} \in E(\mathcal{I})$ by a vertex $v \in V(P \cap Q)$. However, in MSO$_2$ we cannot pick a single vertex from $V(P \cap Q)$ and therefore will represent the edge $\{P, Q\}$ by the set $V(P \cap Q)$. Let

$$\varphi_E(x) := \exists P \in \mathcal{P}\ \exists Q \in \mathcal{Q}\ x \in V(P \cap Q)$$
$$\mathrm{inc}(x, P) := x \in V(P)$$
$$x \sim y := \exists P \in \mathcal{P}\ \exists Q \in \mathcal{Q}\ x, y \in V(P \cap Q)$$

be MSO$_2$-formulas, where we will usually write $\sim(x, y)$ in infix notation. $\sim$ defines an equivalence relation on the set of vertices satisfying $\varphi_E(x)$ and we can represent edges in $\mathcal{I}$ by equivalence classes of $\sim$ in $G$. Hence, $\mathcal{I}$ is isomorphic to the graph $\mathfrak{I} := (V, E, \sigma)$ with vertex set $V := \mathcal{P} \cup \mathcal{Q}$ and edge set $E := \{[x]_\sim : x \in \varphi_E(G)\}$, where a vertex $P \in V$ is incident with an edge $e \in E$ if there is a vertex $v \in e \cap P$ (and hence $e \subseteq P$). $\mathfrak{I}$ is MSO$_2$-definable in $G$, by the formulas $\varphi_E$, inc and $\sim$ with parameters $\mathcal{P}, \mathcal{Q}$ and we can represent variables over elements of $\mathfrak{I}$ by variables ranging over sets of edges in $G$ by enforcing that these are interpreted by a path from $\mathcal{P}$ or $\mathcal{Q}$. Variables $X$ over sets of elements of $\mathfrak{I}$ can be represented in $G$ by pairs $X_P, X_Q$ of variables ranging over sets of edges so that a set $X \subseteq V(\mathfrak{I})$ is represented by the pair of sets $X_P := X \cap \mathcal{P}$ and $X_Q := X \cap \mathcal{Q}$. Finally, sets $F \subseteq E(\mathfrak{I})$ of edges can be represented by sets $F' \subseteq \varphi_E(G)$ closed under $\sim$ so that if $\{P, Q\} \in F$ then $V(P \cap Q) \subseteq F'$. Using this idea we can then say about $\mathfrak{J}$, and hence about $\mathcal{I}$, that $\mathfrak{J}$ is a wall as follows:

1. There are two sets $\mathcal{H}, \mathcal{V} \subseteq E(\mathcal{I})$ of edges, each of which induces a set of pairwise vertex disjoint paths in $\mathcal{I}$ (which we will think of as horizontal and vertical paths in a wall).
2. For all $P \in \mathcal{H}$ and $Q \in \mathcal{V}$, $P \cap Q$ is connected and $V(P \cap Q) \cap V(H) = \emptyset$ for all $H \in (\mathcal{V} \cup \mathcal{H}) \setminus \{P, Q\}$.
3. There is a path $L \in \mathcal{V}$ such that the intersection of $L$ with each $Q \in \mathcal{H}$ contains an endpoint of $Q$ (we think of $L$ as the left-most vertical path in the wall). Once we have $L$, we can give the horizontal paths $P \in \mathcal{H}$ a direction, where we say that $p \in V(P)$ is to *the left* of $p' \in V(P)$, if the subpath of $P$ containing $p'$ and a vertex in $L$ also contains $p$.
4. There is a path $T \in \mathcal{H}$ such that the intersection of $T$ with each $P \in \mathcal{V}$ contains an endpoint of $P$ ($T$ is the top-most horizontal path in the wall). We can now use $T$ to give the vertical paths $P \in \mathcal{V}$ a direction and say that $p \in V(P)$ is *above* $p' \in V(P)$, if the subpath of $P$ containing $p'$ and a vertex in $T$ also contains $p$.
5. For each path $P \in \mathcal{V}$ except $L$ there is a path $P' \in \mathcal{V}$ (the path immediately to the left of $P$) such that for all $Q \in \mathcal{H}$: if $p \in V(P \cap Q)$ and $p' \in V(P' \cap Q)$ are vertices in the intersection of $Q$ and $P$, $P'$ resp., then $p'$ is to the left of $p$ in $Q$ and there is no $S \in \mathcal{H}$ such that any $s \in V(S \cap Q)$ lies in the subpath of $Q$ between $p$ and $p'$.
6. The analogue condition for horizontal paths.

To demonstrate the idea of the MSO$_2$-formalisation we give precise formulas for the set $\mathcal{H}$ in Condition 1. It will be clear that the other conditions can be formalised analogously.

We have to say that there is a set $\mathcal{H} \subseteq E(\mathfrak{J})$ of edges inducing a set of pairwise disjoint paths in $\mathfrak{J}$. To define this in $G$, we first need a formula $\text{Path}(P, Q, H)$ saying that there is a path from $P \in \mathcal{P} \cup \mathcal{Q}$ to $Q \in \mathcal{P} \cup \mathcal{Q}$ using only edges from $H$, where $H$ is a subset of $\varphi_E(G)$, closed under $\sim$, representing edges in $\mathfrak{J}$. The usual way of expressing that two vertices $x, y$ in a graph are connected within a set $H$ of edges is to say that all sets $U$ of vertices which contain $x$ and are closed under the edge relation $H$ also contain $y$. In our case, sets of vertices of $\mathfrak{J}$ are represented by pairs of sets $U_P \subseteq \mathcal{P}$ and $U_Q \subseteq \mathcal{Q}$ consisting of connected components of $\mathcal{P}$ and $\mathcal{Q}$. Hence, the idea above is expressed by the formula $\text{Path}(P, Q, H)$ defined as

$$\forall U_P \subseteq \mathcal{P} \forall U_Q \subseteq \mathcal{Q} \Big(
\begin{aligned}
&\big(P \in U_P \cup U_Q \wedge \\
&\quad \forall X, Y \in \mathcal{P} \cup \mathcal{Q}[U \in U_P \cup U_Q \wedge \exists e(e \in H \wedge \text{inc}(e, X) \wedge \text{inc}(Y))) \\
&\quad \to Y \in U_P \cup U_Q]\big) \\
&\qquad\qquad\qquad\qquad\qquad \to Q \in U_P \cup U_q\Big)
\end{aligned}$$

where we write $X \in U_P \cup U_Q$ to say that $X$ is a component either of $U_P$ or $U_Q$ and $U_P \subseteq \mathcal{P}$ to say that $U_P$ is a set of components of $\mathcal{P}$.

Now, we can say that $\mathcal{H}$ induces a set of pairwise disjoint paths as follows. We first say that every vertex in $\mathcal{H}$ has degree at most 2: $\forall P \in \mathcal{P} \cup \mathcal{Q}(\exists^{\leq 2} f \in \mathcal{H}\, \text{inc}(f, P))$, where $\exists^{\leq 2} f...$ is an abbreviation for: there are at most 2 edges $f$ such that .... To say that $\mathcal{H}$ induces an acyclic graph we say that for all $P \in \mathcal{P} \cup \mathcal{Q}$, if $P$ is incident to an edge $e := \{P, Q\} \in \mathcal{H}$ then there is no path from $P$ to $Q$ in $\mathcal{H} - e$. The latter can be expressed using the formula Path above.

Clearly, if $\mathcal{V}$ and $\mathcal{H}$ satisfy the conditions above, then they generate a wall in $\mathfrak{I}$ and conversely, the disjoint horizontal and vertical paths of a wall satisfy the conditions. Hence, $\mathfrak{I}$ is a wall if such $\mathcal{V}$ and $\mathcal{H}$ exist containing all vertices and edges of $\mathfrak{I}$. Formalising all this gives us a formula which says of $\mathcal{P}, \mathcal{Q}$ that the pair $(\mathcal{P}, \mathcal{Q})$ is a pseudo-wall. Note that so far we have not used the additional path $A$. Hence, if we are not interested in coloured pseudo-walls but simply in pseudo-walls we can use this formula.

We now proceed to define coloured walls and the induced colouring of $\mathcal{I}(\mathcal{P}, \mathcal{Q})$. From the formalisation above we now have sets $\mathcal{H}, \mathcal{V}$ containing the horizontal and vertical paths of the wall as well as two paths $L, T$ giving the top-most row and left-most column. The left-most row gives us an ordering on the top-most row and all we have to do is to define the colours of the vertices on the top-most row from the additional path $A$, which is easily done. Hence, we can write formulas $\varphi_C(P)$, for $C \in \Sigma$ which are true for the vertices in the wall coloured by $C$. Complex pseudo-walls can be defined analogously.

Taken together, we have a formula $\varphi_U(\mathcal{P}, \mathcal{Q}, A)$ which says that $(\mathcal{P}, \mathcal{Q}, A)$ is a coloured pseudo-wall. Here, the sets $\mathcal{P}$ and $\mathcal{Q}$ define the vertices of the pseudo-wall whereas $A$ is an additional parameter used in the formulas. It will be convenient to take the sets $T, L$ defining the top- and left-most row and column as parameters also rather than defining them. Hence, we have a formula $\varphi_U(\mathcal{P}, \mathcal{Q}, A, L, T)$ which says that $(\mathcal{P}, \mathcal{Q}, A)$ is a $\Sigma$-coloured pseudo-wall with left-most column $L$ and top-most row $T$, formulas $\varphi_E(x, \mathcal{P}, \mathcal{Q}, A, L, T)$, $\text{inc}(x, P, \mathcal{P}, \mathcal{Q}, A, L, T)$ and $\sim(x, y, \mathcal{P}, \mathcal{Q}, A, L, T)$ defining the edge relation of the pseudo-wall and formulas $\varphi_B(x, \mathcal{P}, \mathcal{Q}, A, L, T)$ and $\varphi_C(P, \mathcal{P}, \mathcal{Q}, A, L, T)$, where $C \in \Sigma \dot\cup \{R\}$, defining the colours.

All formulas together define, in graphs of large enough tree-width coloured properly, a large wall whose top-most row is labelled by a word over $\Sigma$. Hence, if $\mathcal{C}$ is a class of graphs of unbounded tree-width, closed under colourings, we can define arbitrary large coloured walls in $\mathcal{C}$. We know already that (presumably) MSO-model checking is not fixed-parameter tractable on the class of coloured walls. To prove the main result of this paper we need a way to translate $\text{MSO}_2$-formulas $\varphi$ over walls to $\text{MSO}_2$-formulas $\varphi^*$ over the graphs in which we define the walls. We could do this in an ad-hoc way and modify the formulas $\varphi_U...$ for each given formula $\varphi$. We find it more convenient, though, to treat these modifications uniformly within the framework of interpretations. In the next section we therefore introduce a new form of interpretations which simplifies dealing with the intersection graphs we have to define and which might also be useful elsewhere.

## 5 $\text{MSO}_2 - \text{MSO}_2$-transductions

In this section we introduce a class of interpretations, called $\text{MSO}_2 - \text{MSO}_2$-*transductions*, between classes of graphs which allow us to define one class $\mathcal{B}$ of graphs inside another class $\mathcal{C}$ so that we can translate $\text{MSO}_2$-formulas over $\mathcal{B}$ to $\text{MSO}_2$-formulas saying the same over the graphs in $\mathcal{C}$. To simplify the presentation we will restrict attention to signatures of coloured graphs. Let $\sigma := \{E, B_1, \ldots, B_t, C_1, \ldots, C_s\}$ be a signature of coloured graphs as defined in Section 2. Let $\tau$ be a signature.

**Definition 5.1** (MSO$_2$–MSO$_2$-transduction)**.** Let $\overline{U} := U_1, \ldots, U_k$ and $\overline{X} := X_1, \ldots, X_l$ be tuples of binary relation symbols. An MSO$_2$ − MSO$_2$-*transduction of $\sigma$ in $\tau$ with parameters $\overline{U}, \overline{X}$* is a tuple

$$\Theta := \begin{pmatrix} \varphi_U(U_1, \ldots, U_k, \overline{X}), \\ \left(\varphi_E^{i,j}(x), \text{inc}_E^{i,j}(x, P, Q), \sim^{i,j}(x)\right)_{1 \leq i < j \leq k} \\ \left(\varphi_F^{i,j}(x)\right)_{1 \leq i < j \leq k, F \in \{B_1, \ldots, B_t\}}, \\ \left(\varphi_C(P)^i\right)_{C \in \sigma, 1 \leq i \leq k}, \end{pmatrix}$$

where $P, Q$ are unary second-order variables and $x$ is a first-order variable, such that for all $\tau$-structures $A$ and sets $\overline{U}, \overline{X} \subseteq E(A)$ with $(A, \overline{U}, \overline{X}) \models \varphi_U$:

- $\sim^{i,j}$ defines an equivalence relation on $\varphi_E^{i,j}(A)$
- for all $x \in V(A)$ and $1 \leq i < j \leq k$, if $(A, \overline{U}, \overline{X}) \models \varphi_E^{i,j}(x)$ then there are exactly two sets $P_i \subseteq U_i$ and $P_j \subseteq U_j$ such that $(A, \overline{U}, \overline{X}) \models \text{inc}_E^{i,j}(x, P_i, P_j)$ and if $(A, \overline{U}, \overline{X}) \models x \sim^{i,j} y$ then $(A, \overline{U}, \overline{X}) \models \text{inc}_E^{i,j}(y, P_i, P_j)$
- for all $F \in \{B_1, \ldots, B_t\}$, $\varphi_F(A) \subseteq \varphi_E(A)$.

We abbreviate MSO$_2$−MSO$_2$-transductions of $\sigma$ in $\tau$ as $\sigma$-$\tau$-transductions. Let $\Theta$ be a $\sigma$-$\tau$-transduction. To every $\tau$-structure $A$, $\Theta$ associates a class $\Theta(A)$ of $\sigma$-structures defined as follows. If $U_1, \ldots, U_k, X_1, \ldots, X_l \subseteq E(A)$ are sets of edges such that $(A, \overline{U}, \overline{X}) \models \varphi_U$, then we define the structure $B := \Theta(A, \overline{U}, \overline{X})$ as follows:

- $V(B) := \dot\bigcup_{1 \leq i \leq k} V_i$ where $V_i := \{V \subseteq U_i : V \text{ is a connected component of } U_i\}$
- $E(B) := \dot\bigcup_{1 \leq i < j \leq k} E^{i,j}$ where $E^{i,j} := \{[v]_{\sim^{i,j}} : v \in \varphi_E^{i,j}(A)\}$ and the $E^{i,j}$ are taken to be disjoint.
- an edge $e \in E^{i,j}$ is incident to vertices $P \in V_i$ and $Q \in V_j$ if $A \models \text{inc}^{i,j}(e, P, Q)$ for some $Q \in V_j$ and likewise for $P \in V_j$.
- an edge $e \in E^{i,j}$, for $1 \leq i < j \leq k$, is coloured by $F$, where $F \in \sigma$ is binary, if $A \models \varphi_F^{i,j}(e)$.
- a vertex $P \in V_i$ it coloured by $C \in \sigma$, where $C$ is unary, if $(A, \overline{U}, \overline{X}) \models \varphi_C^i(P)$.

Hence, with every structure $A$ and satisfying assignment $U_1, \ldots, U_k, \overline{X}$ of $\varphi_U$ the transduction $\Theta$ associates structures whose universes consist of the connected components of the $U_i$. Note that, unlike first-order interpretations, MSO$_2$−MSO$_2$-transductions associate with every structure a class of structures and in this sense resemble MSO-transductions as, e.g., studied by Courcelle. The definition of the edge relation may seem to be overly complicated, as we define the edges and their incidence by different formulas and furthermore do it separately for each pair $i, j$. The reason is that we want to use MSO$_2$-formulas over the structures $\Theta(A)$ and hence have to be able to quantify over sets of edges in $B \in \Theta(A)$. As MSO$_2$ does not allow quantification over arbitrary binary relations, we have to encode edges by individual elements of $A$ and then use sets over vertices to encode sets of edges. For classes $\mathcal{A}$ of $\tau$-structures we define
$\Theta(\mathcal{A}) := \{B : B \in \Theta(A) \text{ for some } A \in \mathcal{A}\}$.

$\text{MSO}_2 - \text{MSO}_2$-transductions define a way of transforming one class of coloured graphs into another. On the other hand, they provide a way of translating $\text{MSO}_2$-formulas over $\sigma$-structures into $\text{MSO}_2$-formulas over $\tau$-structures. By induction on the structure of formulas, we define for every formula $\varphi \in \text{MSO}_2[\sigma]$ a $\Theta(\varphi) \in \text{MSO}_2[\tau]$ as follows.

- If $\varphi := \exists F \psi$, where $F$ is a variable ranging over sets of edges, then
$$\varphi^* := \exists (F_{i,j}^*)_{1 \leq i < j \leq k} \big( \bigwedge_{1 \leq i < j \leq k} \forall x (x \in F_{i,j}^* \to \varphi_E^{i,j}(x)) \wedge$$
$$F_{i,j}^* \text{"is closed under } \sim^{i,j}\text{"}) \wedge \psi^*(\overline{F^*}).$$
- If $\varphi := \exists X \psi$, where $X$ is a variable ranging over sets of vertices, then $\varphi^* := \exists X_1^* \ldots X_k^* \big($ "if $P$ is a conn. comp. of $X_i^*$ then $P$ is a conn. comp. of $U_i$" $\wedge$ "$\psi^*(X_1^*, \ldots, X_k^*)\big)$, where $X_i^*$ are fresh variables ranging over sets of edges.
- If $\varphi := \exists x \psi$, where $x$ is first-order, then $\varphi^* := \exists X_1^* \ldots X_k^*$ "exactly one of the $X_i$ is non-empty, say $X_i$" $\wedge$ "$X_i$ contains exactly one conn. comp. $P$ which is a conn. comp. of $U_i$" $\wedge \psi^*(\overline{X})$, where the $X_i$ is a fresh set of variables ranging over sets of vertices.
- Boolean connectives are translated literally, i.e. $(\varphi \vee \psi)* := \varphi^* \vee \psi^*$.
- If $\varphi := Exy$ then $\varphi^*(X_1^*, \ldots, X_k^*, Y_1^*, \ldots, Y_k^*) := \bigvee_{i,j=1}^{k} \big( X_i \neq \varnothing \wedge Y_j \neq \varnothing \wedge \exists e \varphi_E^{i,j}(e) \wedge \text{inc}_E^{i,j}(e, X_i, Y_j) \big)$. The translation for $\varphi(x,y) := B_i xy$ is similar.
- If $\varphi(x) := Cx$, for some unary $C \in \sigma$, then $\varphi^*(X_1^*, \ldots, X_k^*) := \bigvee_{i=1}^{k} \big( X_i^* \neq \varnothing \wedge \varphi_C(X_i^*) \big)$.
- Finally, if $\varphi(x, y) := x = y$ then $\varphi^*(\overline{X^*}, \overline{Y^*}) := \overline{X^*} = \overline{Y^*}$.

Hence, if $\varphi$ is a formula with free variables $F_1, \ldots, F_l, X_1, \ldots, X_s, y_1, \ldots, y_r$, where the $F_i$'s are binary, the $X_i$'s are unary and the $y_i$'s are individual variables, then $\varphi^*$ is a formula with free variables $(F_i)_1^*, \ldots, (F_i)_k^*, (X_i)_1^*, \ldots, (X_i)_k^*,$ and $(Y_i)_1^*, \ldots, (Y_i)_k^*$, where the $(F_i)_j^*$'s are binary and all other unary. In addition, the parameters $\overline{U}, \overline{X}$ of the transduction occur free in $\varphi^*$. We now prove the analogue of the classical interpretation lemma for the case of transductions.

**Lemma 5.2** (interpretation lemma). *Let $A$ be a $\tau$-structure and $\overline{U}, \overline{X} \subseteq E(A)$ be such that $(A, \overline{U}, \overline{X}) \models \varphi_U$. Let $B := \Theta(A, \overline{U}, \overline{X})$. For all $\varphi \in \text{MSO}_2[\sigma]$, $(A, \overline{U}, \overline{X}) \models \Theta(\varphi)$ if, and only if, $B \models \varphi$.*

*Proof.* We prove the following stronger statement by induction on the structure of the formula.

*Claim.* Let $\varphi \in \text{MSO}[\sigma]$ be a formula with free variables $F_1, \ldots, F_l, X_1, \ldots, X_s, y_1, \ldots, y_r$, where the $F_i$'s are binary, the $X_i$'s are unary and the $y_i$'s individual variables.

Then $\varphi^*$ is a formula with free variables $(F_i)_1^*, \ldots, (F_i)_k^*, (X_i)_1^*, \ldots, (X_i)_k^*,$ and $(Y_i)_1^*, \ldots, (Y_i)_k^*$, where the $(F_i)_j^*$'s are binary and all other unary, but also the parameters $\overline{U}, \overline{X}$ of the transduction occur free in $\varphi^*$.

If $F_i \subseteq E(B), X_i \subseteq V(B)$ and $y_i \in V(B)$ is an interpretation of the free variables of $\varphi$ and $\overline{F_i^*}, \overline{X_i^*}, \overline{Y_i^*}$ are interpretations of the variables in $A$ such that for all $i$,

- if $\{P,Q\} \in F_l$ and $P$ is a component of $U_i$ and $Q$ a component of $U_j$ then $(F_l)_{i,j}^*$ contains the element $x$ such that $\mathrm{inc}_E^{i,j}(x,P,Q)$ and conversely, if $(F_l)_{i,j}^*$ contains this element $x$ then $\{P,Q\} \in F_l$.
- if $\{P\} \in X_l$ and $P$ is a component of $U_i$ then $P \in (X_l)_i^*$ and conversely any component $P$ of a $(X_l)_i^*$ occurs in $X_l$
- if $y_l := P$ for some component of $U_i$ then $(Y_l)_I := \{P\}$ and $(Y_l)_j := \varnothing$ for all $j \neq i$

then $(A, \overline{(F_i)^*}, \overline{(X_i)^*}, \overline{(Y_i)^*}) \models \varphi^*$ if, and only if, $(B, \overline{F}, \overline{X}, \overline{y}) \models \varphi$.

The claim is easily proved by induction on the formulas. Clearly, this implies the lemma. □

**Corollary 5.3.** *Let $\varphi \in \mathrm{MSO}_2[\sigma]$ and $\psi := \exists \overline{U} \exists \overline{X} \varphi^* \in \mathrm{MSO}_2[\tau]$.*

1. *For all $\tau$-structures $A$, $A \models \psi$ if, and only if, there is a $B \in \Theta(A)$ such that $B \models \varphi$.*
2. *For all classes $\mathcal{A}$ of $\tau$-structures, $\psi$ is true in a structure $A \in \mathcal{A}$ if, and only if, $\varphi$ is true in a structure $B \in \Theta(\mathcal{A})$.*

This corollary has the obvious consequences, for instance if the $\mathrm{MSO}_2$-theory of $\mathcal{B} := \Theta(\mathcal{A})$ is undecidable then so is the $\mathrm{MSO}_2$-theory of $\mathcal{A}$.

Note that the transductions as defined here are introduced specifically for coloured graphs which is the topic of this paper. To give a more general definition for arbitrary relational structures, it would be interesting to consider *guarded second-order logic* (GSO) as introduced by Grädel, Hirsch and Otto [13]. As the results we obtain in this paper refer most naturally to graphs, we defer a full definition in terms of GSO to the full version of this paper.

## 6 Putting it all together

In this section we prove the main result of this paper. Let $\Sigma := \{C_1, \ldots, C_l\}$, with $l \geq 2$, be a set of colours and $\Gamma := \Sigma \dot{\cup} \{B, R\}$, where $B$ is a binary and $R$ a unary relation symbol.

**Theorem 6.1.** *Let $\mathcal{C}$ be a constructible class of $\Gamma$-coloured graphs closed under colourings.*

1. *If the tree-width of $\mathcal{C}$ is not poly-logarithmically bounded, then $\mathrm{MC}(\mathrm{MSO}_2, \mathcal{C})$ is not in XP and hence not fixed-parameter tractable unless all problems in the polynomial-time hierarchy can be solved in subexponential time.*
2. *If the tree-width of $\mathcal{C}$ is not bounded by $\log^{16} n$, then $\mathrm{MC}(\mathrm{MSO}_2, \mathcal{C})$ is not in XP and hence not fixed-parameter tractable unless $\mathrm{SAT}$ can be solved in sub-exponential time.*

We first observe that the formulas $\varphi_U(\mathcal{P}, \mathcal{Q}, A, L, T), \varphi_E, \mathrm{inc}, \sim, \varphi_B, \varphi_C$ as constructed in Section 4 can be used to define an $\mathrm{MSO}_2 - \mathrm{MSO}_2$-transduction $\Theta$ such that $\Theta(\mathcal{C})$ is the class of coloured walls in graphs $G \in \mathcal{C}$. Here, we take $\overline{U} := \mathcal{P}, \mathcal{Q}$ as the parameters defining the vertex set of the resulting graphs and $\overline{X} := A, L, T$ as additional parameters used in the transduction.

By Lemma 3.13, as $\mathcal{C}$ is rich and closed under colourings, if the tree-width of $\mathcal{C}$ is not bounded by $\log^{8k} n$, for some $k$, then $\Theta(\mathcal{C})$ contains for each $w \in \Sigma^*$ a wall encoding $w$ with power $k$, i.e. there is a $(|w|^k \times |w|^k)$-wall in $\Theta(\mathcal{C})$ whose top-most row is labelled by $w$ from the left.

In particular, as SAT can be solved in time quadratic in the size of the input by a non-deterministic Turing-machine, if we take $k = 2$ then for each CNF formula $w$ of length $m$, $\Theta(\mathcal{C})$ contains a wall of size $m^2 \times m^2$ labelled by $w$.

Now take a formula $\varphi_{\text{CNF}}$ which, on a wall $W$ encoding $w$, checks whether $w$ correctly encodes a CNF-formula and whether the order of $W$ is at least $|w|^2$. This can be done by simulating a non-deterministic Turing machine doing this test. Let $\psi_{\text{CNF}} := \exists \mathcal{PQALT}(\varphi_U \wedge \Theta(\varphi_{\text{CNF}}))$. Finally, let $\varphi$ be the MSO$_2$-sentence from Section 2 which, by simulating an appropriate Turing-machine, is true in a wall of order $|w|^2$ encoding a CNF-formula $w$ if, and only if, $w$ is satisfiable and let $\vartheta := \Theta(\varphi)$. The next lemma follows from the Interpretation Lemma 5.2.

**Lemma 6.2.** *Let $\mathcal{C}$ be a rich class of graphs and $\mathcal{C}_{\text{CNF}} := \{ A \in \mathcal{C} : A \models \psi_{\text{CNF}} \} \subseteq \mathcal{C}$. Then $\mathcal{C}_{\text{CNF}}$ contains for each CNF-formula $w$ a graph $G \in \mathcal{C}$ encoding $w$ with power 2 and conversely each graph $G \in \mathcal{C}_{\text{CNF}}$ encodes a CNF-formula with power 2.*

It follows that if the tree-width of $\mathcal{C}$ is not bounded by $\log^{16} n$, then model-checking $\vartheta := \Theta(\varphi)$ in $\mathcal{C}_{\text{CNF}}$ is equivalent to solving SAT. If in addition $\mathcal{C}$ is constructible then this allows us to formally define a subexponential time reduction from SAT to $\mathcal{C}$.

**Lemma 6.3.** *If $\mathcal{C}$ is a rich and constructible class of $\Gamma$-coloured graphs closed under colourings whose tree-width is not bounded by $\log^{16} n$, then $\text{MC}(\text{MSO}_2, \mathcal{C})$ is not in XP and hence not fixed-parameter tractable unless SAT can be solved in sub-exponential time, i.e. the ETH fails.*

*Proof.* We define the reduction from SAT as follows. Given a CNF-formula $w$, we construct a graph $G \in \mathcal{C}$ such that $G$ encodes $w$ with power 2 and $|G| < 2^{c \cdot |w|^{\frac{1}{y}}}$, for some $y > 1$ and $c > 0$. By the richness condition, such a graph $G$ exists in $\mathcal{C}$ and by the constructability condition it can be constructed in time $|G|^r$, for some fixed $r > 0$, and hence in time $< (2^{|w|^{\frac{1}{y}}})^r = 2^{r \cdot |w|^{\frac{1}{y}}}$.

Now suppose $\text{MC}(\text{MSO}_2, \mathcal{C})$ was in XP, i.e. given a graph $G \in \mathcal{C}$ and $\varphi \in \text{MSO}_2$, $G \models \varphi$ could be decided in time $|G|^{f(|\varphi|)}$, for some computable function $f$. Hence, we could decide if $G \models \vartheta$, where $\vartheta$ is the formula defined above, in time $|G|^{f(|\vartheta|)} < 2^{f(|\vartheta|) \cdot |w|^{\frac{1}{y}}}$.

Taken together, we could decide if $w$ is satisfiable in time $< 2^{(r + f(|\vartheta|)) \cdot |w|^{\frac{1}{y}}}$, for fixed $r, y > 1$ and a fixed formula $\vartheta$. Hence, SAT would be decidable in sub-exponential time. $\square$

The same argument shows the following corollary.

**Corollary 6.4.** *If $\mathcal{C}$ is a rich and constructible class of $\Gamma$-coloured graphs closed under colourings whose tree-width is not bounded by $\log^{8 \cdot k} n$ and $\mathcal{L}$ is a problem than can be* *decided by a non-deterministic Turing-machine in time $n^k$, then* $\mathrm{MC}(\mathrm{MSO}_2, \mathcal{C})$ *is not in XP and hence not fixed-parameter tractable unless $\mathcal{L}$ can be solved in sub-exponential time.*

Clearly, Lemma 6.3 and Corollary 6.4 together imply Theorem 6.1. The extension to the polynomial time hierarchy follows as we can simulate alternating Turing-machines with bounded number of alternations in $\mathrm{MSO}_2$ in the same way as non-deterministic Turing-machines.